\documentclass[12pt]{article}
\usepackage{array}
\usepackage{epsfig}

\usepackage{amssymb}

\makeatletter
\def\alt{\mathrel{\mathpalette\gl@align<}}
\def\agt{\mathrel{\mathpalette\gl@align>}}
\def\gl@align#1#2{\lower.6ex\vbox{\baselineskip\z@skip\lineskip\z@
\ialign{$\m@th#1\hfil##\hfil$\crcr#2\crcr\sim\crcr}}} \makeatother

\def\bwt{\begin{widetext}}
\def\ewt{\end{widetext}}
\def\be{\begin{equation}}
\def\ee{\end{equation}}
\def\bea{\begin{eqnarray}}
\def\eea{\end{eqnarray}}
\def\bean{\begin{eqnarray*}}
\def\eean{\end{eqnarray*}}
\def\bary{\begin{array}}
\def\eary{\end{array}}
\def\bit{\begin{itemize}}
\def\eit{\end{itemize}}

\def\su5u1{SU(5) \times U(1)}
\def\fsu5u1{SU(5) \times U(1)'}
\def\so10{SO(10)}
\def\sq20{SO(10) \times SO(10)}

\begin{document}
\begin{flushright}
{\tt hep-ph/0608181}\\
BA-06-16, RUNHETC-06-19\\
\end{flushright}
\vspace*{1.0cm}
\begin{center}
{\baselineskip 25pt \Large{\bf
Non-Canonical MSSM, Unification, \\ And New Particles At The LHC  \\[2.5mm]
}}

\vspace{1cm}

{\large Ilia Gogoladze$^a$\footnote {On a leave of absence from:
Andronikashvili Institute of Physics, GAS, 380077 Tbilisi, Georgia.
\\ \hspace*{0.5cm} email: {\tt ilia@physics.udel.edu}}, Tianjun Li$^{b,c}$\footnote
{email: {\tt tjli@physics.rutgers.edu}},
V. N. {\c S}eno$\breve{\textrm{g}}$uz$^d$\footnote {email: {\tt nefer@udel.edu}}
  and Qaisar
Shafi$^d$\footnote {email: {\tt shafi@bartol.udel.edu}} }
\vspace{.5cm}

{\small {\it $^a$Department of Physics and Astronomy, University of
Delaware, Newark, DE 19716, USA \\
$^b$Department of Physics and Astronomy, Rutgers University,
Piscataway, NJ 08854, USA\\
$^c$Institute of Theoretical Physics, Chinese Academy of Sciences,
 Beijing 100080, P. R. China \\
$^d$ Bartol Research Institute, Department of Physics and Astronomy,
University of Delaware, Newark, DE 19716, USA }}


\date{\today}

\vspace{1.0cm} {\bf Abstract}

\end{center}

\baselineskip 16pt

We consider non-canonical embeddings of the MSSM
in high-dimensional orbifold GUTs based on the gauge symmetry
$SU(N)$, $N=5,6,7,8$. The hypercharge normalization factor $k_Y$ can either
have unique non-canonical values, such as $23/21$ in a six-dimensional
$SU(7)$ model, or may lie in a (continuous) interval. Gauge coupling unification 
and gauge-Yukawa unification can be realized in these models by introducing
new particles with masses in the TeV range which may be found at the LHC. In one such example
there exist color singlet fractionally charged states.

\thispagestyle{empty}

\bigskip
\newpage

\addtocounter{page}{-1}

\section{Introduction}
High-dimensional orbifold grand unified theories (GUTs) \cite{Orbifold,Li:2001tx} 
provide the elegant solutions to the well-known problems encountered in four-dimensional
(4D) GUTs such as $SU(5)$ and $SO(10)$, especially the
doublet-triplet splitting problem and the proton decay problem.
The non-supersymmetric version has, in
particular, been exploited to show that unification of the standard model
(SM) gauge couplings can be realized with a non-canonical embedding of
$U(1)_Y$, the hypercharge component of the SM gauge group \cite{Barger:2005gn}. The couplings
unify at $M_{\rm GUT}\simeq4\times10^{16}$ GeV, which is also the scale at which the
4D ${\cal N} = 1$ supersymmetry (SUSY) is broken, without introducing additional new
particles. This approach has been taken a step further along two different
directions. In \cite{Gogoladze:2006ps} it was shown that by implementing additional gauge-Yukawa 
unification, the SM Higgs mass can be predicted. The mass turns out to 
be $135\pm6$ ($144\pm4$) GeV with gauge-top (bottom/tau) Yukawa unification.
This is encouraging because it is different from the 
prediction of $\lesssim130$ GeV in the minimal supersymmetric standard model (MSSM). 
In \cite{Gogoladze:2006qp} these ideas were extended to the case
of split supersymmetry, with similar predictions
for the Higgs mass.


The orbifold scenario for the GUT breakings
assume the supersymmetric GUT models exist in high dimensions
and are broken to 4D ${\cal N}=1$ supersymmetric standard like
models for the zero modes due to the discrete symmetries on
the extra space manifolds~\cite{Orbifold,Li:2001tx}. The zero
modes can be identified with the low-energy SM fermions and Higgs
fields, allowing gauge-Higgs unification~\cite{Antoniadis:2001cv}
and gauge-Yukawa unification~\cite{GY}.
For the canonical $U(1)_Y$ normalization,
the unification of the gauge couplings, top and bottom quark
Yukawa couplings, and $\tau$ lepton Yukawa coupling can be
realized in the 6D orbifold $SU(8)$ and $SU(9)$
models, and cannot be obtained in the orbifold $SU(N)$
models with $N < 8$. Therefore, it is interesting to construct
the minimal orbifold $SU(N)$ model with gauge-Yukawa
unification.

In this paper, we show that the minimal model with the
unification of the gauge couplings and third-family
Yukawa couplings is the 6D orbifold $SU(7)$ model
with non-canonical $U(1)_Y$ normalization $k_Y=23/21$
where $k_Y$ is defined in Eqs. (1) and (2). Moreover,
we construct the 7D $SU(8)$ models with
gauge-Yukawa unification and $k_Y > 23/21$. And for
completeness, we consider the 6D orbifold
$SU(5)$ and $SU(6)$ models with gauge-fermion and
gauge-fermion-Higgs unification first
as warm up exercise.

The 4D gauge group in these models is $SU(3)_C\times
SU(2)_L\times U(1)_Y$ accompanied by one or 
several extra $U(1)$ factors assumed to be broken at $M_{\rm GUT}$.
We define the unified gauge couplings at the GUT scale ($M_{\rm GUT}$) as 
\begin{eqnarray}
g_1^2  ~=~ g_2^2 ~=~ g_3^2~,~\,
\end{eqnarray}
where 
\begin{eqnarray}
g_1^2 \equiv k_Y g_Y^2~,~\,
\end{eqnarray}
where $k_Y$ is the $U(1)_Y$ normalization factor, and the
$g_Y$, $g_2$, and $g_3$ are the gauge couplings for
$U(1)_Y$, $SU(2)_L$, and $SU(3)_C$ gauge groups, respectively.
For the canonical $U(1)_Y$ normalization, we have $k_Y=5/3$.

For orbifold GUTs where all of the SM fermions and Higgs fields
are placed on a 3-brane at an orbifold fixed point, we can have 
any positive normalization for $U(1)_Y$, {\it i.~e.}, 
$k_Y$ is an arbitrary positive real number. However, in this case charge quantization cannot
be realized. We wish to consider the more interesting orbifold GUTs in which at
least one of the SM fermions and Higgs fields arise from
the zero modes of the bulk vector multiplet and their $U(1)_Y$
charges can be determined. The charge quantization can be
achieved due to the gauge invariance of Yukawa couplings
and anomaly free conditions. In the orbifold models we consider, 
$k_Y$ is then either uniquely determined to have a non-canonical
value or lies in a continuous interval. For the latter case $k_Y=5/3$ is possible,
but there is no apparent reason why this value would be realized. 

Since the three SM gauge couplings unify quite nicely with the canonical
hypercharge normalization, it can be argued that we should simply discard the
models which do not predict $k_Y=5/3$. However, unification in MSSM with
$k_Y=5/3$ may well be accidental, and as the example of non-supersymmetric
unification shows there are different possibilities.  In this paper we assume a
non-canonical hypercharge normalization as the models under consideration generally
predict. We then discuss how gauge coupling unification and gauge-Yukawa unification can 
be obtained by adding a minimal set of vector-like particles to the MSSM spectrum. 
It is certainly our hope that these vector-like particles will be found at the
Large Hadron Collider (LHC).

The paper is organized as follows. In sections 2 and 3 we consider $SU(5)$ and
$SU(6)$ models.  In the $SU(5)$ model the only zero mode that can be introduced
in the bulk is a quark doublet and $k_Y$ is predicted to be $1/15$.  The model
can be extended to $SU(6)$, with $k_Y\ge1/15$. We construct two $SU(6)$ models
with gauge-top and gauge-bottom Yukawa coupling unification, with $k_Y=2/3$ and
$1/3$ respectively. We discuss $SU(7)$ and $SU(8)$ models in sections 4 and 5.
We can have gauge-Yukawa unification for the third family in an $SU(7)$ model,
with $k_Y=23/21$.  This model can be extended to $SU(8)$, with $k_Y\ge23/21$.
Sections 6 and 7 concern gauge coupling unification and gauge-Yukawa
unification with new particles in these models.  We briefly remark on the Higgs
mass in section 8 and conclude in section 9. Some details of the 6D and 7D
orbifold models are provided in the two appendices. 

\section{$SU(5)$ Models}

We consider a 6D ${\cal N} = (1, 1)$ supersymmetric  
$SU(5)$ gauge theory compactified on the
orbifold $M^4\times T^2/Z_6$ (for some details see
Appendix A). The ${\cal N} = (1, 1)$ supersymmetry in 6D
has 16 supercharges and
 corresponds to ${\cal N}=4$ supersymmetry in 4D,
and thus only the gauge multiplet can be introduced in the bulk.  This
multiplet can be decomposed under the 4D
 ${\cal N}=1$ supersymmetry into a vector
multiplet $V$ and three chiral multiplets $\Sigma_1$, $\Sigma_2$, and 
$\Sigma_3$ in the adjoint representation, where the fifth and sixth 
components of the gauge field, $A_5$ and $A_6$, are contained in 
the lowest component of $\Sigma_1$.

To break the $SU(5)$ gauge symmetry, we choose the following
$5\times 5$ matrix representation for $R_{\Gamma_T}$,
\begin{eqnarray}
R_{\Gamma_T} &=& {\rm diag} \left(+1, +1, +1,
 \omega^{n_1}, \omega^{n_1} \right)~,~\,
 \label{SU5-R-GT}
\end{eqnarray}
where $w=e^{{\rm i}\pi/3} $ and $n_1 \not= 0$.
 Then, we obtain\footnote{Suppose $G$ is a Lie group and $H$ is a subgoup of $G$,
we denote the commutant of $H$ in $G$ as $G/H$, {\it i. e.},
\begin{equation}
G/H \equiv \{g \in G|gh=hg, ~{\rm for ~any} ~ h \in H\}~.~\,\nonumber
\end{equation}}
\begin{eqnarray}
 SU(5)/R_{\Gamma_T} ~=~ SU(3)_C\times SU(2)_L\times U(1)_Y .~\,
\end{eqnarray}
So,  for the zero modes, the 6D  ${\cal N} = (1, 1)$
supersymmetric $SU(5)$ gauge symmetry  is broken down to  4D
${\cal N}=1$ supersymmetric $SU(3)_C\times
SU(2)_L\times U(1)_Y $ gauge symmetry~\cite{Li:2001tx}.

We define the generator for $U(1)_Y$ as follows:
\begin{eqnarray}
T_{U(1)_{Y}} &\equiv& {1\over 30} ~{\rm diag}\left(2, 2, 2, -3, -3 \right) 
~.~\, 
\label{SU5-GTT1Xa} 
\end{eqnarray}
Because ${\rm tr} [T_{U(1)_{Y}}^2]=1/30$, we obtain $k_Y=1/15$.

Under $SU(3)_C\times SU(2)_L\times U(1)_Y$, the adjoint representation $\mathbf{24}$ of $SU(5)$ decomposes 
 as
\begin{equation}
\mathbf{24} = \left(
\begin{array}{cc}
\mathbf{(8,1)}_{Q00} & \mathbf{(3, \bar 2)}_{Q12} \\
 \mathbf{(\bar 3,  2)}_{Q21} & \mathbf{(1,3)}_{Q00}
\end{array}
\right) +  \mathbf{(1,1)}_{Q_{00}}\, ,
\label{24arj}
\end{equation}
where  the last term
$\mathbf{(1,1)}_{Q_{00}}$ denotes the gauge field
associated with $U(1)_Y$.
The subscripts $Qij$, with $Qij=-Qji$, denote the
charges under  $U(1)_Y$, and 
\begin{eqnarray}
Q00~=~\mathbf{0}~~,~Q12=\mathbf{{1\over 6}}~.~\,
\label{SU5-Qij}
\end{eqnarray}

The $Z_6$ transformation properties for the decomposed components
of $V$, $\Sigma_1$, $\Sigma_2$,  and $\Sigma_3$ are given by the
first $2\times 2$ submatrices in  
Eqs. (\ref{T2trans1})--(\ref{T2trans4}) in Appendix A. 
We choose
\begin{eqnarray}
k~=~1 ~,~~ n_1~=~5~,~\,
\label{SU5-I-N-numberA}
\end{eqnarray}
where $k$ is given in Eqs. (\ref{S2trans-6D}) and (\ref{S3trans-6D})
in Appendix A.
There are no zero modes from the chiral
multiplets $\Sigma_2$ and $\Sigma_3$,
and only one zero mode, $\mathbf{(3, \bar 2)}_{Q12}$,
from the chiral multiplet $\Sigma_1$, which can be identified as
the third-family quark doublet $Q_3$.
The remaining MSSM matter fields and the two MSSM Higgs doublets
can be put on the 3-brane at $z=0$, where only the SM gauge symmetry
is preserved.


\section{$SU(6)$ Models}
 
For the $SU(6)$ models where at least one of the SM fermions
and Higgs fields arise from the zero modes of  the chiral
multiplets $\Sigma_1$, $\Sigma_2$ and $\Sigma_3$, we can
show that the minimal normalization $k_Y$ for $U(1)_Y$
is 1/15, and the corresponding zero mode is quark doublet 
because it has the smallest $U(1)_Y$ quantum number.
Moreover, we can only have the gauge-top or 
gauge-bottom quark Yukawa coupling unification, and
we cannot obtain the right-handed leptons from the
zero modes of bulk vector multiplet.

In the following subsections, we present three $SU(6)$ models.
In the first, the third family quark doublet $Q_3$
is the only zero mode from the bulk vector multiplet,
and $k_Y$ is an arbitrary positive real number that
is larger than or equal to 1/15.
In the second and third $SU(6)$ models, we have
gauge-top and gauge-bottom quark Yukawa coupling
unification, respectively. We consider 7D ${\cal N} =1$ supersymmetric  
$SU(6)$ compactified on the
orbifold $M^4\times T^2/Z_6 \times S^1/Z_2$ (for some details see
Appendix B), and 6D ${\cal N} = (1, 1)$ supersymmetric  
$SU(6)$ compactified on the
orbifold $M^4\times T^2/Z_6$.
The compactification process yields 4D
${\cal N}=1$ supersymmetric $SU(3)_C\times
SU(2)_L\times U(1)_Y \times U(1)_{\alpha}$.

The generators for
$U(1)_Y \times U(1)_{\alpha}$ are defined as follows:
\begin{eqnarray}
T_{U(1)_{Y}} &\equiv& {1\over 30} ~{\rm diag}\left(2, 2, 2, -3, -3,
 0 \right) + a ~{\rm diag}\left(1, 1,
1, 1, 1, -5 \right),~\, \label{SU5-GTT1X} \nonumber
\\ 
T_{U(1)_{\alpha}} &\equiv& ~{\rm diag}\left(2, 2, 2, -3, -3, 0 \right) 
- {{1}\over {30a}} ~{\rm diag}\left(1, 1,
1, 1, 1, -5 \right),~\, \label{SU5-GTT2X} 
\end{eqnarray}
where $a$ is a real number.
Because ${\rm tr} [T_{U(1)_{Y}}^2]=1/30 + 30a^2$, we obtain 
\begin{eqnarray}
k_Y~=~{{1}\over {15}} + 60 a^2 ~\ge~ {{1}\over {15}} ~.~\,
\end{eqnarray}

The adjoint representation $\mathbf{35}$ of $SU(6)$ is decomposed under
 $SU(3)_C\times SU(2)_L\times U(1)_Y \times U(1)_{\alpha}$ 
as
\begin{equation}
\mathbf{35} = \left(
\begin{array}{ccc}
\mathbf{(8,1)}_{Q00} & \mathbf{(3, \bar 2)}_{Q12}
& \mathbf{(3, 1)}_{Q13}  \\
 \mathbf{(\bar 3,  2)}_{Q21} & \mathbf{(1,3)}_{Q00}
& \mathbf{(1, 2)}_{Q23} \\
\mathbf{(\bar 3, 1)}_{Q31} & \mathbf{(1, \bar 2)}_{Q32}
& \mathbf{(1, 1)}_{Q00} 
\end{array}
\right) +  \mathbf{(1,1)}_{Q_{00}}\, ,
\label{48arja}
\end{equation}
where  $\mathbf{(1,1)}_{Q00}$ in the
third diagonal entry of the matrix and the last term
$\mathbf{(1,1)}_{Q_{00}}$ denote  gauge fields
associated with $U(1)_Y \times U(1)_{\alpha} $.
The subscripts $Qij$, with $Qij=-Qji$, are the
charges under  $U(1)_Y \times U(1)_{\alpha} $.
The subscript $Q00~=~(\mathbf{0}, \mathbf{0})$, and the
other subscripts $Qij$ with $i\not= j$ are
\begin{eqnarray}
 Q12=(\mathbf{{1\over 6}}, \mathbf{5})~,~
Q13=(\mathbf{{1\over 15}+6a},  \mathbf{2-{1\over {5a}}})~,~
Q23=(\mathbf{-{1\over 10}+6a},  \mathbf{-3-{1\over {5a}}})~.~
\label{SU6-Qij}
\end{eqnarray}
We will consider the following three  models. 

\subsection{ $SU(6)$ Model I}

Here
 the third-family quark doublet $Q_3$ is 
the only zero mode from the bulk vector multiplet, $a$ is an arbitrary real number, and
 we have 
\begin{eqnarray}
k_Y ~\ge~ {{1}\over {15}} ~.~\,
\end{eqnarray}

To project out all the unwanted components in
the chiral multiplets, we consider
the 7D ${\cal N} =1$ supersymmetric  
$SU(6)$.
The ${\cal N}=1$ supersymmetry in 7D has 16 supercharges corresponding
to ${\cal N}=4$ supersymmetry in 4D, and only the
gauge supermultiplet can be introduced in the bulk.  This multiplet
can be decomposed under  4D
 ${\cal N}=1$ supersymmetry into a gauge vector
multiplet $V$ and three chiral multiplets $\Sigma_1$, $\Sigma_2$,
and $\Sigma_3$  all in the adjoint representation, where the fifth
and sixth components of the gauge field, $A_5$ and $A_6$, are
contained in the lowest component of $\Sigma_1$, and the seventh
component of the gauge field $A_7$ is contained in the lowest
component of $\Sigma_2$.

To break the $SU(6)$ gauge symmetry, we choose the 
following $6\times 6$ matrix
representations for $R_{\Gamma_T}$ and $R_{\Gamma_S}$
\begin{eqnarray}
R_{\Gamma_T} &=& {\rm diag} \left(+1, +1, +1, 
 \omega^{n_1}, \omega^{n_1}, \omega^{n_2} \right)~,~\,
\end{eqnarray}
\begin{eqnarray}
R_{\Gamma_S} &=& {\rm diag} \left( +1, +1, +1, +1, +1, +1\right)~,~\,
\end{eqnarray}
where $n_1 \not= n_2 \not= 0$.

Then, we obtain
\begin{eqnarray}
 SU(6)/R_{\Gamma_T} ~=~ SU(3)\times SU(2) \times U(1)_Y \times U(1)_{\alpha} ~,~\,
\end{eqnarray}
\begin{eqnarray}
SU(6)/R_{\Gamma_S} ~=~ SU(6) ~,~\,
\end{eqnarray}
\begin{eqnarray}
 SU(6)/\{R_{\Gamma_T} \cup R_{\Gamma_S}\}
~=~SU(3)\times SU(2) \times U(1)_Y \times U(1)_{\alpha}~.~\,
\end{eqnarray}
Note that $R_{\Gamma_S}$ only breaks the additional supersymmetry.

The $Z_6\times Z_2$ transformation properties for the decomposed components
of $V$, $\Sigma_1$, $\Sigma_2$,  and $\Sigma_3$ are the $3\times 3$
submatrices  in
 Eqs. (\ref{SU8-GTBT-trans-1})--(\ref{SU8-GTBT-trans-4}) in Appendix B
where the third and fourth rows and columns are crossed out.
We choose
\begin{eqnarray}
n_1~=~5~,~~n_2~=~2~~{\rm or}~~3~.~\,
\label{SU6-N-numberB}
\end{eqnarray}
Then, we obtain that
there is no zero mode from the chiral
multiplets $\Sigma_2$ and $\Sigma_3$,
and only one zero mode, $\mathbf{(3, \bar 2)}_{Q12}$,
from the chiral multiplet $\Sigma_1$, which can be identified with
the third-family quark doublet $Q_3$.

\subsection{ $SU(6)$ Model II and $SU(6)$ Model III}

In this subsection, we will construct $SU(6)$ models
with gauge-top and gauge-bottom quark Yukawa coupling
unification. We consider 6D ${\cal N} = (1, 1)$ supersymmetric  
$SU(6)$ compactified on the orbifold $M^4\times T^2/Z_6$.
To break the $SU(6)$ gauge symmetry, we choose the following
$6\times 6$ matrix representation for $R_{\Gamma_T}$ 
\begin{eqnarray}
R_{\Gamma_T} &=& {\rm diag} \left(+1, +1, +1,
 \omega^{n_1}, \omega^{n_1},  \omega^{n_2} \right)~,~\,
 \label{SU6-TB-GT}
\end{eqnarray}
where $n_1 \not= n_2 \not= 0$.
 Then, we obtain
\begin{eqnarray}
 SU(6)/R_{\Gamma_T} ~=~ SU(3)_C\times SU(2)_L\times U(1)_Y 
\times U(1)_{\alpha}.~\,
\end{eqnarray}

The $Z_6$ transformation properties for the decomposed components
of $V$, $\Sigma_1$, $\Sigma_2$,  and $\Sigma_3$ are given by the
first $3\times 3$ submatrices in  Eqs. (\ref{T2trans1})--(\ref{T2trans4})
in Appendix A.  We choose
\begin{eqnarray}
k~=~1 ~,~~ n_1~=~5~,~~n_2~=~2~,~\,
\label{SU6-TB-numberA}
\end{eqnarray}
and consider the following two models: \\

(A) $SU(6)$ Model II (gauge-top quark Yukawa coupling unification)\\

With
\begin{eqnarray}
a~=~ {1\over {10}}~,~\,
\end{eqnarray}
we have
\begin{eqnarray}
k_Y~=~ {2\over 3}~.~\,
\end{eqnarray}
The zero modes from  the chiral
multiplets $\Sigma_1$, $\Sigma_2$ and $\Sigma_3$ are presented in
Table \ref{Spectrum-GT-SU6}. We can identify them as 
the third-family quark doublet, the right-handed top
quark, and the MSSM Higgs doublets.

\renewcommand{\arraystretch}{1.4}
\begin{table}[ht]
\begin{center}
\begin{tabular}{|c|c|}
\hline
Chiral Fields & Zero Modes  \\
\hline\hline
$\Sigma_1$ & $Q_3$:~ $\mathbf{(3, \bar 2)}_{Q12}$ \\
\hline 
$\Sigma_2$ & $t^c$:~ $\mathbf{(\bar 3 , 1)}_{Q31}$ \\
\hline
$\Sigma_3$ &  ~$H_u$:~ $\mathbf{(1, 2)}_{Q23}$;
~$H_d$:~ $\mathbf{(1, \bar 2)}_{Q32}$ \\
\hline
\end{tabular}
\end{center}
\vspace{-0.3cm} \caption{\small Zero modes from  the chiral
multiplets $\Sigma_1$, $\Sigma_2$ and $\Sigma_3$ in
$SU(6)$ (Model II).}
\label{Spectrum-GT-SU6}
\end{table}

From the trilinear term in the 6D bulk action, we obtain
the top quark Yukawa term
\begin{eqnarray}
 \int d^6 x \left[ \int d^2 \theta \ g_6  Q_3 t^c H_u  + h.c.\right]~.~\,
\end{eqnarray}
Thus, at  $M_{\rm GUT}$, we have
\begin{eqnarray}
g_1=g_2=g_3=y_t=g_6/\sqrt{V}~,~\,
\end{eqnarray}
where $y_t$ is the top quark Yukawa coupling, 
and $V$ is the physical volume of extra dimensions. \\

(B) $SU(6)$ Model III (gauge-bottom quark Yukawa coupling unification)\\

For this case we set
\begin{eqnarray}
a~=~ -{1\over {15}}~,~\,
\end{eqnarray}
in which case
\begin{eqnarray}
k_Y~=~ {1\over 3}~.~\,
\end{eqnarray}
The zero modes arise from the chiral
multiplets $\Sigma_1$, $\Sigma_2$ and $\Sigma_3$, 
and are presented in
Table \ref{Spectrum-GB-SU6}. We can identify them as 
the third-family quark doublet, the right-handed bottom
quark, and the MSSM Higgs doublets.

\renewcommand{\arraystretch}{1.4}
\begin{table}[ht]
\begin{center}
\begin{tabular}{|c|c|}
\hline
Chiral Fields & Zero Modes  \\
\hline\hline
$\Sigma_1$ & $Q_3$:~ $\mathbf{(3, \bar 2)}_{Q12}$ \\
\hline 
$\Sigma_2$ & $b^c$:~ $\mathbf{(\bar 3 , 1)}_{Q31}$ \\
\hline
$\Sigma_3$ &  ~$H_d$:~ $\mathbf{(1, 2)}_{Q23}$;
~$H_u$:~ $\mathbf{(1, \bar 2)}_{Q32}$ \\
\hline
\end{tabular}
\end{center}
\vspace{-0.3cm} \caption{\small Zero modes from  the chiral
multiplets $\Sigma_1$, $\Sigma_2$ and $\Sigma_3$ in
 $SU(6)$ (Model III).}
\label{Spectrum-GB-SU6}
\end{table}

From the trilinear term in the 6D bulk action, we obtain
the bottom quark Yukawa term
\begin{eqnarray}
 \int d^6 x \left[ \int d^2 \theta \ g_6  Q_3 b^c H_d  + h.c.\right]~.~\,
\end{eqnarray}
Thus, at  $M_{\rm GUT}$, we have
\begin{eqnarray}
g_1=g_2=g_3=y_b=g_6/\sqrt{V}~,~\,
\end{eqnarray}
where $y_b$ is the bottom quark Yukawa coupling.


\section{$SU(7)$ Models}

As we discussed above, to achieve
gauge-fermion-Higgs unification, the minimal gauge group is $SU(7)$,
with $U(1)_Y$ normalization $k_Y=23/21$ which is uniquely determined.
This can be seen as follows.
The $U(1)_Y$ generator in $SU(7)$ belongs to 
its Cartan subalgebra, and can be parame\-trized as
\begin{eqnarray}
T_{U(1)_{Y}} \equiv {\rm diag}\left(r_3, r_3, r_3, r_2, r_2, r_1, r'_1 \right)~.~
\end{eqnarray}  
The traceless condition yields 
\begin{eqnarray}
3 r_3 + 2 r_2 + r_1 + r'_1 ~=~0~,~\,
\end{eqnarray}  
and gauge-fermion-Higgs unification requires that
\begin{eqnarray}
r_3-r_2~=~{1\over 6}~,~~r_3-r_1~=~{2\over 3}~,~~r_3-r'_1~=~-{1\over 3}~.~\,
\end{eqnarray}  
Thus, we have the unique solution
\begin{eqnarray}
r_3~=~ {{2}\over {21}}~,~~r_2~=~ -{{1}\over {14}}~,
~~r_1~=~ -{{4}\over {7}}~,~r'_1~=~ {{3}\over {7}}~,~\,
\end{eqnarray}  
for which  ${\rm tr} [T_{U(1)_{Y}}^2]=23/42$.
With a canonical normalization ${\rm tr}[T_i^2]=1/2$ of non-abelian
generators, we obtain $k_Y = 23/21$.

We consider a 6D ${\cal N} = (1, 1)$ supersymmetric  
$SU(7)$ gauge theory compactified on the
orbifold $M^4\times T^2/Z_6$ (for some details see
Appendix A). 
To break $SU(7)$, we select the following
$7\times 7$ matrix representation for $R_{\Gamma_T}$ 
\begin{eqnarray}
R_{\Gamma_T} &=& {\rm diag} \left(+1, +1, +1,
 \omega^{n_1}, \omega^{n_1}, \omega^{n_2}, \omega^{n_3} \right)~,~\,
 \label{SU7-R-GT}
\end{eqnarray}
where $n_1 \not= n_2 \not= n_3 \not= 0$.
 Thus,
\begin{eqnarray}
 SU(7)/R_{\Gamma_T} ~=~ SU(3)_C\times SU(2)_L\times U(1)_Y \times U(1)_{\beta} \times
U(1)_{\gamma}~.~\,
\end{eqnarray}
So,  for the zero modes, the 6D  ${\cal N} = (1, 1)$
supersymmetric $SU(7)$ gauge symmetry  is broken down to  4D
${\cal N}=1$ supersymmetric $SU(3)_C\times
SU(2)_L\times U(1)_Y \times U(1)_{\beta} \times U(1)_{\gamma} $
gauge symmetry~\cite{Li:2001tx}.
We assume that the two additional  $U(1)$ 
symmetries can be spontaneously broken at $M_{\rm GUT}$ by the usual Higgs mechanism. It is conceivable
that these two symmetries can play some useful role as flavor
symmetries \cite{FN}, but we will not pursue this any further here.

We define the generators for the
$U(1)_Y \times U(1)_{\beta}\times U(1)_{\gamma}$ gauge symmetry
 as follows
\begin{eqnarray}
T_{U(1)_{Y}} &\equiv& {1\over 42} ~{\rm diag}\left(4, 4, 4, -3, -3,
-24, 18 \right) ~,~\, \label{SU7-GTT1X} \nonumber
\\ 
 T_{U(1)_{\beta}}
&\equiv& {\rm diag}\left(1, 1, 1, 0, 0, -1, -2 \right),~\, 
\label{SU7-GTT3X} \nonumber \\
 T_{U(1)_{\gamma}}
&\equiv& {\rm diag}\left(3, 3, 3, -8, -8, 5, 2\right).~\,
\label{SU7-GTT1A}
\end{eqnarray}

The $SU(7)$ adjoint representation $\mathbf{48}$ decomposes under
the $SU(3)_C\times SU(2)_L\times U(1)_Y \times U(1)_{\beta} \times
U(1)_{\gamma}$ gauge symmetry as
\begin{equation}
\mathbf{48} = \left(
\begin{array}{cccc}
\mathbf{(8,1)}_{Q00} & \mathbf{(3, \bar 2)}_{Q12}
& \mathbf{(3, 1)}_{Q13} & \mathbf{(3,1)}_{Q14} \\
 \mathbf{(\bar 3,  2)}_{Q21} & \mathbf{(1,3)}_{Q00}
& \mathbf{(1, 2)}_{Q23} & \mathbf{(1, 2)}_{Q24} \\
\mathbf{(\bar 3, 1)}_{Q31} & \mathbf{(1, \bar 2)}_{Q32}
& \mathbf{(1, 1)}_{Q00} & \mathbf{(1, 1)}_{Q34} \\
\mathbf{(\bar 3, 1)}_{Q41} & \mathbf{(1, \bar 2)}_{Q42}
& \mathbf{(1, 1)}_{Q43} & \mathbf{(1, 1)}_{Q00}
\end{array}
\right) +  \mathbf{(1,1)}_{Q_{00}}\, ,
\label{48arj}
\end{equation}
where  $\mathbf{(1,1)}_{Q00}$ in the
third and fourth diagonal entries of the matrix and the last term
$\mathbf{(1,1)}_{Q_{00}}$ denote the gauge fields
associated with $U(1)_Y \times U(1)_{\beta} \times U(1)_{\gamma} $.
The subscripts $Qij$, which are anti-symmetric ($Qij=-Qji$), are the
charges under  $U(1)_Y \times U(1)_{\beta} \times U(1)_{\gamma}$.
The subscript $Q00~=~(\mathbf{0}, \mathbf{0}, \mathbf{0})$, and the
other subscripts $Qij$ with $i\not= j$ are
\begin{eqnarray}
&& Q12=(\mathbf{{1\over 6}}, \mathbf{1}, \mathbf{11})~,~
Q13=(\mathbf{2\over 3},  \mathbf{2}, \mathbf{-2})~,~\nonumber \\
 && Q14=(\mathbf{-{1\over 3}},  \mathbf{3}, \mathbf{1})~,~
Q23=(\mathbf{1\over 2},  \mathbf{1}, \mathbf{-13}) ~,~\nonumber \\
 && Q24=(\mathbf{-{1\over 2}},  \mathbf{2}, \mathbf{-10}) ~,~
Q34=(\mathbf{-1},  \mathbf{1}, \mathbf{3}) ~.~\,
\label{SU7-GTT-Qij}
\end{eqnarray}

The $Z_6$ transformation properties for the decomposed components
of $V$, $\Sigma_1$, $\Sigma_2$,  and $\Sigma_3$ are given by
 Eqs. (\ref{T2trans1})--(\ref{T2trans4}). 
We will consider two concrete models.

\subsection{ $SU(7)$ Model I}

We choose
\begin{eqnarray}
k~=~1 ~,~~ n_1~=~4~,~~n_2~=~1~,~~ n_3~=~2~,~\,
\label{SU7-I-N-numberA}
\end{eqnarray}
where $k$ is given in Eqs. (\ref{S2trans-6D}) and (\ref{S3trans-6D})
in Appendix A. The zero modes from  the chiral
multiplets $\Sigma_1$, $\Sigma_2$ and $\Sigma_3$ are presented in
Table \ref{Spectrum-GTBT-SU7-I}. We can identify them as 
the third-family SM fermions, and 
one pair of Higgs doublets. Interestingly, we do not
have any exotic particle from the zero modes of the chiral
multiplets.

\renewcommand{\arraystretch}{1.4}
\begin{table}[ht]
\begin{center}
\begin{tabular}{|c|c|}
\hline
Chiral Fields & Zero Modes  \\
\hline\hline
$\Sigma_1$ & $t^c$: $\mathbf{(\bar 3, 1)}_{Q31}$;
~$\tau^c$:~ $\mathbf{(1, 1)}_{Q43}$ \\
\hline 
$\Sigma_2$ & $Q_3$:~ $\mathbf{(3, \bar 2)}_{Q12}$;
~$H_d$:~ $\mathbf{(1,  2)}_{Q24}$;
~$b^c$:~ $\mathbf{(\bar 3 , 1)}_{Q41}$ \\
\hline
$\Sigma_3$ & $H_u$:~ $\mathbf{(1, 2)}_{Q23}$;
~$L_3$:~ $\mathbf{(1, \bar 2)}_{Q32}$ \\
\hline
\end{tabular}
\end{center}
\vspace{-0.3cm} \caption{\small Zero modes from  the chiral
multiplets $\Sigma_1$, $\Sigma_2$ and $\Sigma_3$  in $SU(7)$ (Model I).}
\label{Spectrum-GTBT-SU7-I}
\end{table}

From the trilinear term in the 6D bulk action, we obtain
the top quark and tau lepton Yukawa terms
\begin{eqnarray}
 \int d^6 x \left[ \int d^2 \theta \ g_7 \left( Q_3 t^c H_u 
+ L_3 \tau^c H_d \right) + h.c.\right]~.~\,
\end{eqnarray}
Thus, at  $M_{\rm GUT}$, we have
\begin{eqnarray}
g_1=g_2=g_3=y_t=y_{\tau}=g_7/\sqrt{V}~,~\,
\end{eqnarray}
where $y_{\tau}$ is the tau lepton Yukawa coupling.
However, we do not have the bottom quark
Yukawa term from 6D bulk action.

\subsection{ $SU(7)$ Model II}

We choose
\begin{eqnarray}
k~=~1 ~,~~ n_1~=~4~,~~n_2~=~1~,~~ n_3~=~3~.~\,
\label{SU7-I-N-numberB}
\end{eqnarray}
The zero modes from  the chiral
multiplets $\Sigma_1$, $\Sigma_2$ and $\Sigma_3$ are given in
Table \ref{Spectrum-GTBT-SU7-II}. We can identify them as 
the third-family SM fermions,
the MSSM Higgs doublets, and an exotic (left-handed singlet) quark $b_X$.

\renewcommand{\arraystretch}{1.4}
\begin{table}[ht]
\begin{center}
\begin{tabular}{|c|c|}
\hline
Chiral Fields & Zero Modes  \\
\hline\hline
$\Sigma_1$ & $H_d$:~ $\mathbf{(1,  2)}_{Q24}$;
~$t^c$: $\mathbf{(\bar 3, 1)}_{Q31}$ \\
\hline
$\Sigma_2$ & $Q_3$:~ $\mathbf{(3, \bar 2)}_{Q12}$;
~$\tau^c$:~ $\mathbf{(1, 1)}_{Q43}$ \\
\hline 
$\Sigma_3$ & ~$H_u$:~ $\mathbf{(1, 2)}_{Q23}$;
~$L_3$:~ $\mathbf{(1, \bar 2)}_{Q32}$;
~$b^c$:~ $\mathbf{(\bar 3 , 1)}_{Q41}$;
~$b_X$:~ $\mathbf{(3 , 1)}_{Q14}$ \\
\hline
\end{tabular}
\end{center}
\vspace{-0.3cm} \caption{\small Zero modes from  the chiral
multiplets $\Sigma_1$, $\Sigma_2$ and $\Sigma_3$ in 
the $SU(7)$ Model II.}
\label{Spectrum-GTBT-SU7-II}
\end{table}

From the trilinear term in the 6D bulk action, we obtain
the top quark, bottom quark, and tau lepton Yukawa terms
\begin{eqnarray}
 \int d^6 x \left[ \int d^2 \theta \ g_7 \left( Q_3 t^c H_u + Q_3 b^c H_d
+ L_3 \tau^c H_d \right) + h.c.\right].~\,
\end{eqnarray}
Thus, at  $M_{\rm GUT}$, we have
\begin{eqnarray}
g_1=g_2=g_3=y_t=y_b=y_{\tau}=g_7/\sqrt{V}~.~\,
\end{eqnarray}
Thus, we have unification of the SM gauge couplings and the
third-family SM fermion Yukawa couplings.

We can give GUT-scale mass to 
the exotic quark $b_X$ by
introducing an additional exotic quark ${\bar b}_X$
with quantum number $\mathbf{(\bar 3,1)}_{QX}$
on the observable 3-brane at $z=0$,
where $QX=(\mathbf{{1\over 3}, \mathbf{-3}, \mathbf{0}})$.
Suppose we introduce one pair of SM singlets $S'$ and $\overline{S'}$ 
with charges $\mathbf{1}$ and
$\mathbf{-1}$ respectively whose VEVs break $U(1)_{\gamma}$ at $M_{\rm GUT}$.
 The exotic quarks $b_X$ and 
${\bar b}_X$ can pair up and acquire $M_{\rm GUT}$ mass
via the brane-localized superpotential term $ \overline{S'} b_X {\bar b}_X$.

\section{$SU(8)$ Models}

We are unable to construct orbifold models of gauge-fermion-Higgs unification
with $k_Y < 23/21$. To construct models with
 $k_Y \ge 23/21$, we consider a 7D ${\cal N}=1$ supersymmetric  
$SU(8)$ gauge theory compactified on the
orbifold $M^4\times T^2/Z_6 \times S^1/Z_2$ (for some details see
Appendix B). To break the $SU(8)$ gauge symmetry, we choose the 
following $8\times 8$ matrix
representations for $R_{\Gamma_T}$ and $R_{\Gamma_S}$
\begin{eqnarray}
R_{\Gamma_T} &=& {\rm diag} \left(+1, +1, +1, 
 \omega^{n_1}, \omega^{n_1}, \omega^{n_1}, +1, \omega^{n_2} \right)~,~\,
\end{eqnarray}
\begin{eqnarray}
R_{\Gamma_S} &=& {\rm diag} \left( +1, +1, +1, +1, +1, -1, -1, +1\right)~,~\,
\end{eqnarray}
where $n_1 \not= n_2 \not= 0$.
We obtain
\begin{eqnarray}
 SU(8)/R_{\Gamma_T} ~=~ SU(4)\times SU(3) \times U(1)^2 ~,~\,
\end{eqnarray}
\begin{eqnarray}
SU(8)/R_{\Gamma_S} ~=~ SU(6)\times SU(2) \times U(1) ~,~\,
\end{eqnarray}
\begin{eqnarray}
 SU(8)/\{R_{\Gamma_T} \cup R_{\Gamma_S}\}
~=~SU(3)_C \times SU(2)_L\times U(1)_Y \times U(1)_{\alpha}
\times U(1)_{\beta} \times U(1)_{\gamma}~.~\,
\end{eqnarray}
Thus, for the zero modes, the 7D 
${\cal N} = 1 $ supersymmetric $SU(8)$ gauge symmetry is broken 
down to a 4D ${\cal N}=1$ supersymmetric
$SU(3)_C\times SU(2)_L\times U(1)_Y \times U(1)_{\alpha}
\times U(1)_{\beta} \times U(1)_{\gamma}$ gauge symmetry~\cite{Li:2001tx}.

We define the generators for the
$U(1)_Y \times U(1)_{\alpha} \times U(1)_{\beta}\times U(1)_{\gamma}$ gauge symmetry
 as follows:
\begin{eqnarray}
T_{U(1)_{Y}} &\equiv& {1\over 42} ~{\rm diag}\left(4, 4, 4, -3, -3,
-24, 18, 0 \right) + a ~{\rm diag}\left(1, 1,
1, 1, 1, 1, 1, -7 \right),~\, \label{SU8-GTT1X} \nonumber
\\ 
T_{U(1)_{\alpha}} &\equiv& ~{\rm diag}\left(4, 4, 4, -3, -3,
-24, 18, 0 \right) - {{23}\over {56a}} ~{\rm diag}\left(1, 1,
1, 1, 1, 1, 1, -7 \right),~\, \label{SU8-GTT2X} \nonumber
\\
 T_{U(1)_{\beta}}
&\equiv& {\rm diag}\left(1, 1, 1, 0, 0, -1, -2, 0 \right),~\, 
\label{SU8-GTT3X} \nonumber \\
 T_{U(1)_{\gamma}}
&\equiv& {\rm diag}\left(3, 3, 3, -8, -8, 5, 2, 0 \right),~\,
\label{SU8-GTT1A}
\end{eqnarray}
where $a$ is a real number.
Because ${\rm tr} [T_{U(1)_{Y}}^2]=23/42 + 56a^2$, we obtain 
\begin{eqnarray}
k_Y~=~{{23}\over {21}} +112 a^2 ~\ge~ {{23}\over {21}} ~.~\,
\end{eqnarray}
Incidentally, for the canonical $U(1)_Y$ normalization ($k_Y=5/3$), we have
$a=1/14$, and $U(1)_Y$ coincides with $U(1)_Y$ in
the Pati-Salam or Pati-Salam like models when we break $SU(8)$
 down to  $SU(4)_C\times SU(2)_L\times SU(2)_R\times U(1)^2$ or
$SU(3)_C\times SU(2)_L\times U(1)_{B-L} \times U(1)_{I_{3R}}\times U(1)^2$ by orbifold projections.

The $SU(8)$ adjoint representation $\mathbf{63}$
decomposes under  
$SU(3)_C\times SU(2)_L\times U(1)_Y \times U(1)_{\alpha}
\times U(1)_{\beta} \times U(1)_{\gamma}$ gauge symmetry as:
\begin{equation}
\mathbf{63} = \left(
\begin{array}{ccccc}
\mathbf{(8,1)}_{Q00} & \mathbf{(3, \bar 2)}_{Q12} 
& \mathbf{(3, 1)}_{Q13} & \mathbf{(3,1)}_{Q14}
 & \mathbf{(3,1)}_{Q15} \\
 \mathbf{(\bar 3,  2)}_{Q21} & \mathbf{(1,3)}_{Q00}
& \mathbf{(1, 2)}_{Q23} & \mathbf{(1, 2)}_{Q24}
& \mathbf{(1, 2)}_{Q25} \\
\mathbf{(\bar 3, 1)}_{Q31} & \mathbf{(1, \bar 2)}_{Q32}
& \mathbf{(1, 1)}_{Q00} & \mathbf{(1, 1)}_{Q34}
& \mathbf{(1, 1)}_{Q35} \\
\mathbf{(\bar 3, 1)}_{Q41} & \mathbf{(1, \bar 2)}_{Q42}
& \mathbf{(1, 1)}_{Q43} & \mathbf{(1, 1)}_{Q00}
& \mathbf{(1, 1)}_{Q45} \\
\mathbf{(\bar 3, 1)}_{Q51} & \mathbf{(1, \bar 2)}_{Q52}
& \mathbf{(1, 1)}_{Q53} & \mathbf{(1, 1)}_{Q54}
& \mathbf{(1, 1)}_{Q00} 
\end{array}
\right) +  \mathbf{(1,1)}_{Q00}~,~\,
\label{SU8-GTBT}
\end{equation}
where   $\mathbf{(1,1)}_{Q00}$ in the third, fourth and fifth
diagonal entries of the matrix, and the last term $\mathbf{(1,1)}_{Q00}$ 
denote the gauge fields for  $U(1)_Y \times U(1)_{\alpha}
\times U(1)_{\beta} \times U(1)_{\gamma}$. Moreover, 
the subscripts $Qij$, with $Qij=-Qji$, are the charges under
 $U(1)_Y \times U(1)_{\alpha}
\times U(1)_{\beta} \times U(1)_{\gamma}$.
The subscript $Q00~=~ (\mathbf{0}, \mathbf{0}, \mathbf{0}, \mathbf{0})$, 
and the other subscripts $Qij$ with $i\not= j$ are
\begin{eqnarray}
&& Q12=(\mathbf{{1\over 6}}, \mathbf{7}, \mathbf{1}, \mathbf{11})~,~
Q13=(\mathbf{2\over 3}, \mathbf{28}, \mathbf{2}, \mathbf{-2})~,~\nonumber \\
 && Q14=(\mathbf{-{1\over 3}}, \mathbf{-14}, \mathbf{3}, \mathbf{1})~,~
Q23=(\mathbf{1\over 2}, \mathbf{21}, \mathbf{1}, \mathbf{-13}) ~,~\nonumber \\
 && Q24=(\mathbf{-{1\over 2}}, \mathbf{-21}, \mathbf{2}, \mathbf{-10}) ~,~
Q34=(\mathbf{-1}, \mathbf{-42}, \mathbf{1}, \mathbf{3}) ~,~\nonumber \\
 && Q15=\left (\mathbf{{2\over {21}}+8a},
\mathbf{4-{{23}\over {7a}}}, \mathbf{1}, \mathbf{3} \right )~,\nonumber \\
 && Q25=\left (\mathbf{-{1\over {14}}+8a},
\mathbf{-3-{{23}\over {7a}}}, \mathbf{0}, \mathbf{-8} \right )~,\nonumber \\
 && Q35=\left (\mathbf{-{4\over {7}}+8a},
\mathbf{-24-{{23}\over {7a}}}, \mathbf{-1}, \mathbf{5} \right )~,\nonumber \\
 && Q45=\left (\mathbf{{3\over {7}}+8a},
\mathbf{18-{{23}\over {7a}}}, \mathbf{-2}, \mathbf{2} \right )~.
\label{SU8-GTT-Qij}
\end{eqnarray}

The $Z_6\times Z_2$ transformation properties for the decomposed components
of $V$, $\Sigma_1$, $\Sigma_2$,  and $\Sigma_3$ are given by
 Eqs. (\ref{SU8-GTBT-trans-1})--(\ref{SU8-GTBT-trans-4}) in Appendix B.
And we choose
\begin{eqnarray}
n_1~=~5~,~~n_2~=~2~~{\rm or}~~3~.~\,
\label{GTT-N-numberB}
\end{eqnarray}
The zero modes from  the chiral
multiplets $\Sigma_1$, $\Sigma_2$ and $\Sigma_3$
 are presented in
Table \ref{Spectrum-GTBT-7D}. We can identify them as 
the third-family SM fermions,
the MSSM Higgs doublets, and the exotic quark $b_X$.

\renewcommand{\arraystretch}{1.4}
\begin{table}[ht]
\begin{center}
\begin{tabular}{|c|c|}
\hline
Chiral Fields & Zero Modes  \\
\hline\hline
$\Sigma_1$ & $Q_3$:~ $\mathbf{(3, \bar 2)}_{Q12}$;
~$\tau^c$:~ $\mathbf{(1, 1)}_{Q43}$ \\
\hline $\Sigma_2$ & ~$H_u$:~ $\mathbf{(1, 2)}_{Q23}$;
~$L_3$:~ $\mathbf{(1, \bar 2)}_{Q32}$;
~$b^c$:~ $\mathbf{(\bar 3 , 1)}_{Q41}$;
~$b_X$:~ $\mathbf{(3 , 1)}_{Q14}$ \\
\hline
$\Sigma_3$ & $H_d$:~ $\mathbf{(1,  2)}_{Q24}$;
~$t^c$: $\mathbf{(\bar 3, 1)}_{Q31}$ \\
\hline
\end{tabular}
\end{center}
\vspace{-0.3cm} \caption{\small Zero modes from  the chiral
multiplets $\Sigma_1$, $\Sigma_2$ and $\Sigma_3$ in
the $SU(8)$ model.}
\label{Spectrum-GTBT-7D}
\end{table}

\noindent From the trilinear term in the 7D bulk action, we obtain
the top quark, bottom quark, and tau lepton Yukawa terms
\begin{eqnarray}
 \int d^7 x \left[ \int d^2 \theta \ g_8 \left( Q_3 t^c H_u + Q_3 b^c H_d
+ L_3 \tau^c H_d \right) + h.c.\right].~\,
\end{eqnarray}
Thus, at  $M_{\rm GUT}$, we have
\begin{eqnarray}
g_1=g_2=g_3=y_t=y_b=y_{\tau}=g_8/\sqrt{V}~.~\,
\end{eqnarray}

\section{New Particles and Gauge Coupling Unification} \label{su6}
For non-canonical $U(1)_Y$ normalization, it is necessary to introduce new particles 
to achieve unification. Here, as an example, we consider restoring 
gauge coupling unification by adding a minimal set of 
vector-like particles with SM quantum numbers. 
These particles can be put on the 3-brane at $z=0$, and their masses can be
the order of the weak scale due to the Giudice-Masiero mechanism \cite{Giudice:1988yz}.

We denote these particles as $u_x$ and so on, where $u_x$ stands for the vector-like pair with
the same quantum numbers as these for $u+u^c$. 
Although we employ two-loop renormalization group equations
(RGEs) for the gauge couplings
in the numerical calculations, for the discussions below we will consider one-loop $\beta$-coefficients
which, for the MSSM and vector-like particles, are as follows:
\begin{eqnarray} 
b_{\rm MSSM}=\left(\frac{11}{k_Y},1,-3\right),&\quad& b_{Q_x}=\left(\frac{1}{3k_Y},3,2\right),\nonumber \\
b_{u_x}=\left(\frac{8}{3k_Y},0,1\right),&\quad& b_{d_x}=\left(\frac{2}{3k_Y},0,1\right),\nonumber \\
b_{L_x}=\left(\frac{1}{k_Y},1,0\right),&\quad& b_{e_x}=\left(\frac{2}{k_Y},0,0\right)\,.
\label{eq0}
\end{eqnarray}
From the one-loop RGEs, it is straightforward to obtain the following relations:
{\setlength\arraycolsep{2pt}
\begin{eqnarray}
\log\frac{M_{\rm GUT}}{m_Z}&=&\frac{2\pi(\alpha_s^{-1}\alpha-s^2_W)}{\alpha(b_3-b_2)}\,,\label{eq1}\\
\alpha_s^{-1}&=&\left[s^2_W+\frac{1-(1+k_Y)s^2_W}{k_Y}\left(\frac{b_3-b_2}{b_1-b_2}\right)\right]\alpha^{-1}\,,\label{eq2}\\
\alpha_{\rm GUT}&=&\frac{k_Y\alpha(b_1-b_2)}{k_Y s^2_W b_1-(1-s^2_W)b_2}\,,\label{eq3}
\end{eqnarray}}
\noindent where $s_W$ stands for $\sin\theta_W$, and $\alpha$ and $\alpha_s$ are 
the electromagnetic and strong couplings at $m_Z$. From Eq. (\ref{eq0}), we see that $b_3-b_2$ is an
integer. For the GUT scale to be smaller than the Planck scale and large enough to
avoid the bounds on proton decay, Eq. (\ref{eq1}) requires the contribution $(b_3-b_2)_x$ 
from vector-like particles to vanish, assuming the latter have masses close to
the weak scale. From Eq. (\ref{eq2}), the range of $(b_1-b_2)_x$ 
allowing gauge coupling unification can be obtained depending on the value of $k_Y$. 
Also, $\alpha_{\rm GUT}\ll1$ is required for perturbative unification.

Simple examples that satisfy the above conditions are as
follows. For $k_Y=1/15$ as in the $SU(5)$ model, gauge
coupling unification can be restored by adding two sets
of $L_x+u_x$. Unification can also be restored by adding
$L_x+u_x+2e_x$ or by adding $4e_x$. For $k_Y=1/3$ as in
the $SU(6)$ model with gauge-bottom quark Yukawa coupling
unification, one can again add two sets of $L_x+u_x$,
or $3e_x$. And for $k_Y=2/3$ as in the $SU(6)$ model with
gauge-top quark Yukawa coupling unification, one can add
$L_x+u_x+e_x$ or $3(L_x+d_x)+e_x$. Finally, for $k_Y=23/21$
as in the $SU(7)$ model with the unification of the gauge
couplings and third-family Yukawa couplings, one can
add $L_x+u_x$. Because such additional vector-like
particles can be observed at the LHC and ILC, we can
distinguish these models with these future experiments.

\section{New Particles and Gauge-Yukawa Unification}
In this section we probe gauge-Yukawa unification following the analysis in Ref. \cite{Gogoladze:2003pp} 
(see also Ref. \cite{Tobe:2003bc} for details and references).
In our analysis, we use a dimensional reduction ($\overline{\rm DR}$)
renormalization scheme, which is known to be consistent with SUSY.
${\overline{\rm{DR}}}$ Yukawa couplings ($y_{t,b,\tau}$)
and gauge couplings ($g_i$) in the MSSM at Z-boson mass scale
are written as follows:
\begin{eqnarray}
{y}_t(m_Z) &=&
\frac{\sqrt{2}\bar{m}_t^{\rm{MSSM}}(m_Z)}{\bar{v}(m_Z) \sin\beta}
=\frac{\sqrt{2} \bar{m}_t^{\rm{SM}}(m_Z)}{\bar{v}(m_Z) \sin\beta}
\left(1+\delta_t\right),\\
{y}_{b,\tau}(m_Z) &=&
\frac{\sqrt{2}\bar{m}_{b,\tau}^{\rm{MSSM}}(m_Z)}{\bar{v}(m_Z) \cos\beta}
=\frac{\sqrt{2}\bar{m}_{b,\tau}^{\rm{SM}}(m_Z)}{\bar{v}(m_Z) \cos\beta}
\left(1+\delta_{b,\tau}\right),\\
{g}_i(m_Z)&=&
\bar{g}_i^{\rm{SM}}(m_Z)\left(1+\delta_{g_i}\right),~~(i=1-3)
\end{eqnarray}
where $\bar{m}_i^{\rm{SM}}$ and $\bar{g}_i^{\rm{SM}}$ are
$\overline{\rm{DR}}$ quantities defined in the SM, and
$\bar{v}$ and $\tan\beta$ are $\overline{\rm DR}$ values in the MSSM.
They are determined following the analysis in Ref.~\cite{Tobe:2003bc}.
We adopt top pole mass ($m_t=172.5$ GeV) \cite{unknown:2006qt}, tau pole mass
($m_\tau=1777$ MeV), $\overline{\rm{MS}}$
bottom mass ($\bar{m}^{\rm MS}_b(\bar{m}^{\rm MS}_b)
=4.26$ GeV), and $\alpha^{\rm MS}_s(m_Z)=0.119$ \cite{Eidelman:2004wy}.
The quantities $\delta_{t,b,\tau,g_i}$ represent SUSY threshold
corrections. In our analysis, we treat them as free parameters
without specifying any particular SUSY breaking scenario.
When all parameters $\delta_{t,b,\tau,g_i}$ are specified, all
$\overline{\rm{DR}}$ couplings in the MSSM are determined at $m_Z$. Then
we use the two-loop RGEs for the MSSM gauge couplings
and the one-loop RGEs for the Yukawa couplings in order to study the unification of 
couplings at the GUT scale.

In order to study the gauge-Yukawa unification,
we look for a region where top, bottom and
tau Yukawa couplings are unified ($y_t=y_b=y_\tau\equiv y_G$) at the GUT
scale. We define the GUT scale ($M_G$) as a scale where
$g_1(M_G)=g_2(M_G)\equiv g_G$. In our analysis, we allow the possibility that  the strong
gauge coupling is not exactly unified, {\it i.~e.},
$g_3(M_G)^2/g_G^2=1+\epsilon_3$
where $\epsilon_3$ can be a few \%. This mismatch $\epsilon_3$ from
exact unification can be due to the GUT-scale threshold
corrections to the unified gauge coupling.

\begin{figure}[t]
\centering
\includegraphics[angle=0, width=15.5cm]{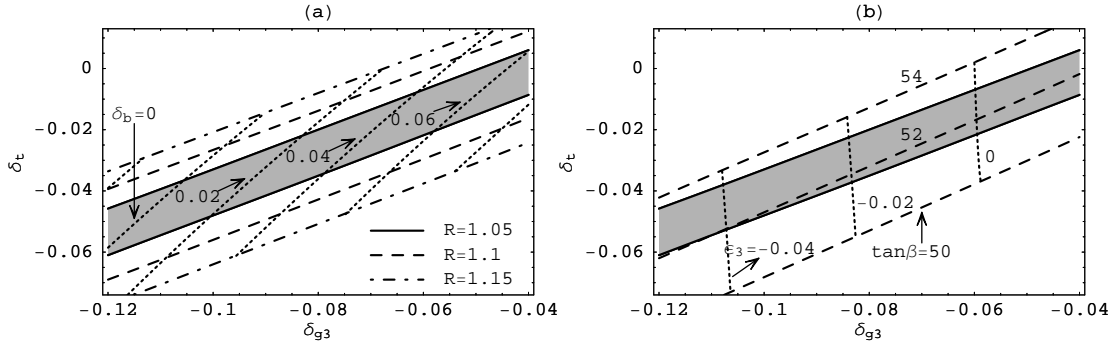} 
\vspace{-.5cm} 
\caption{Parameter space satisfying the gauge-Yukawa unification.
Contours of $\delta_b$ (dotted lines in Fig.~(a)), $\tan\beta$ (dashed lines
in Fig.~(b)) and $\epsilon_3$ (dotted lines in Fig.~(b)) are shown as a
function of $\delta_t$ and $\delta_{g_3}$, required for Yukawa unification
($y_t=y_b=y_\tau$). After finding the region for the Yukawa unification,
contours of a parameter $R$ (defined in text) are plotted in
Fig.~(a). The shaded regions represent a region where the gauge-Yukawa
unification is  achieved within $5\%$ level ($R \leq 1.05$).
Here we have fixed $\delta_\tau=0.02$,
$\delta_{g_1}=-0.006$ and $\delta_{g_2}=-0.02$.} \label{53}
\end{figure}

First, we review gauge-Yukawa unification for the canonical case
$k_Y=5/3$. In Fig.~\ref{53}, contours of $\delta_b$ (dotted
lines in Fig.~(a)), $\tan\beta$ (dashed lines in Fig.~(b)) and
$\epsilon_3$ (dotted lines in Fig.~(b))
are shown as a function of $\delta_t$ and $\delta_{g_3}$,
which are required for the Yukawa unification at the GUT scale.
In order to fix $\delta_{g_{1,2}}$, we assume that all SUSY mass parameters
which contribute to $\delta_{g_{1,2}}$ are equal to $500$ GeV
($\delta_{g_1}=-0.006$ and $\delta_{g_2}=-0.02$).
As shown in Fig.~\ref{53}, 
$\tan\beta$ should be about $52$, and the value of
$\delta_b$ should be a few \%, 
which is much smaller than one naively expected in large $\tan\beta$
case. Small values of $\delta_b$ significantly constrains the superpartner
spectrum, as discussed in Refs.~\cite{Blazek:2001sb,Tobe:2003bc}. On
the other hand, $\delta_t$ is in the expected range (see Ref. \cite{Pierce:1996zz}).

After requiring Yukawa unification, we calculate a parameter $R$
defined as follows:
\begin{eqnarray}
R&\equiv& \frac{{\rm max}(y_G,g_G)}
{{\rm min}(y_G,g_G)}\,.
\end{eqnarray}
In the shaded regions of Fig.~\ref{53}, gauge-Yukawa
unification is realized within $5\%$ level ($R \leq 1.05$), while allowing
$\epsilon_3$ to be a few \%.

\begin{figure}[t] 
\centering
\includegraphics[angle=0, width=15.5cm]{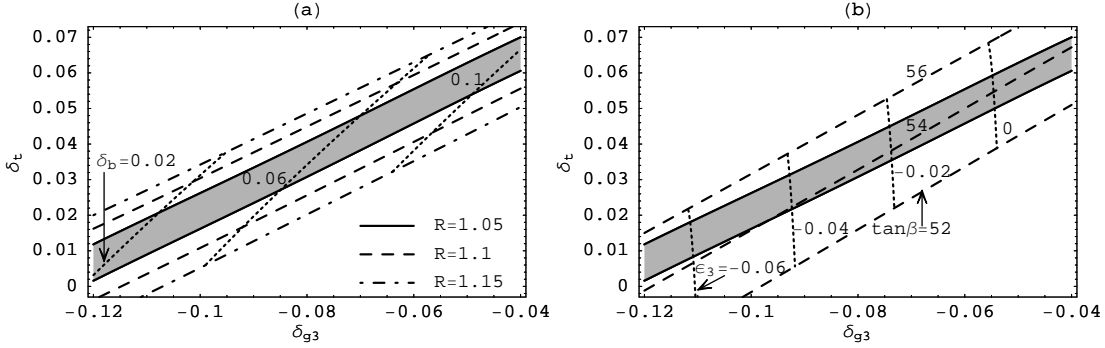} 
\vspace{-.5cm} 
\caption{Same as Fig. 1, but for $k_Y=23/21$ with one set of $L_x+u_x$ added at $M=300$ GeV.} \label{2321ul}
\end{figure}

Next, we take $k_Y=23/21$ as predicted
by the $SU(7)$ model, and give examples as how gauge-Yukawa unification might
be realized. Gauge coupling unification can be restored by adding 
vector-like particles with SM quantum numbers, as in section \ref{su6}. 
A simple example for $k_Y=23/21$ is adding one set of $L_x+u_x$.
However, as shown in Fig. \ref{2321ul}, Yukawa unification then requires
$\delta_t$ shifted up 0.06 compared to Fig. \ref{53}, which is not compatible
with the SUSY threshold corrections in most of the parameter space.

Note that $\delta_t$ can be modified if mixing in the top quark sector is allowed.
We then have the Yukawa and mass terms
\begin{equation}
y_t Q' H_u u^{\prime c} + y'Q'H_u u_x^{\prime c} +M u'_x u_{x}^{\prime c}~,~\,
\end{equation}
where the primes denote weak eigenstates. Diagonalizing the mass matrix, we obtain
\begin{equation}
\frac{y_t}{y_{t0}}=\left(\frac{2}{1+\xi^2+x^2-\sqrt{(1+\xi^2+x^2)^2-4x^2}}\right)^{1/2}\,.
\end{equation}
Here the notation is as follows: $y_{t0}$ is the value without mixing, $x\equiv M/m_t$,
and $\xi\equiv y'/y_t$. Experimentally, $M\lesssim200$ GeV is excluded \cite{Acosta:2002ju}.
As an example we take $M=300$ GeV. Precision electroweak data (more precisely the bounds on the oblique parameter T)
then requires the extended CKM parameter $V_{xb}\lesssim0.4$ \cite{Aguilar-Saavedra:2002kr}. 
This constraint corresponds to $\xi\lesssim0.5$ and a downward shift in $\delta_t$ of $\lesssim0.06$. 

\begin{figure}[t] 
\centering
\includegraphics[angle=0, width=15.5cm]{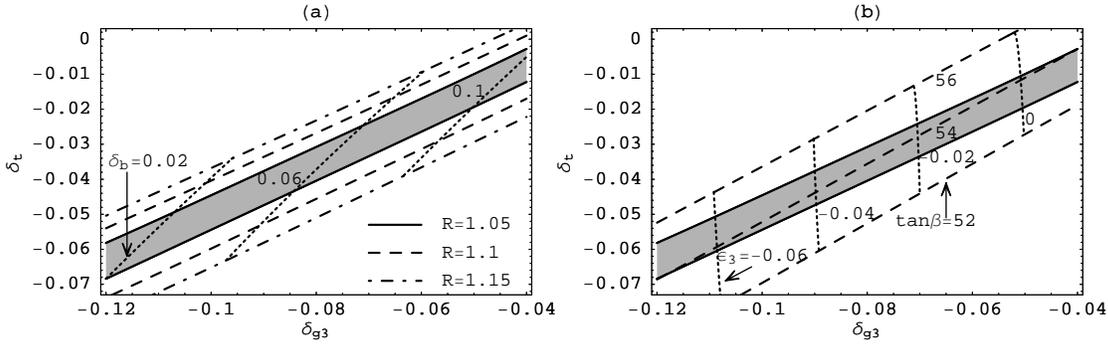} 
\vspace{-.5cm} 
\caption{Same as Fig. 1, but for $k_Y=23/21$ with one set of $L_x+d_x+e_x$ added at $M=300$ GeV.
The Yukawa coupling $y_1$ is assumed negligible, while $y_2$ is taken to be $0.7$ at $M$, corresponding
to $\simeq1.5$ at the GUT scale.} \label{2321dle}
\end{figure}

A similar example is adding one set of $L_x+d_x+e_x$. Gauge-Yukawa unification is then obtained
essentially with the same parameters as above, since the $\beta$-coefficients are identical
at one loop. $\delta_t$ in this case can be modified even assuming no mixing,
due to the new Yukawa couplings $y_1 L_x H_d e^c_x + y_2 L^c_x H_u e_x$. Shifting
$\delta_t$ down appreciably requires no or a weak $y_1$ coupling and a strong $y_2$ coupling,
and a numerical example is provided in Fig. \ref{2321dle}. 
 
\begin{figure}[t] 
\centering
\includegraphics[angle=0, width=15.5cm]{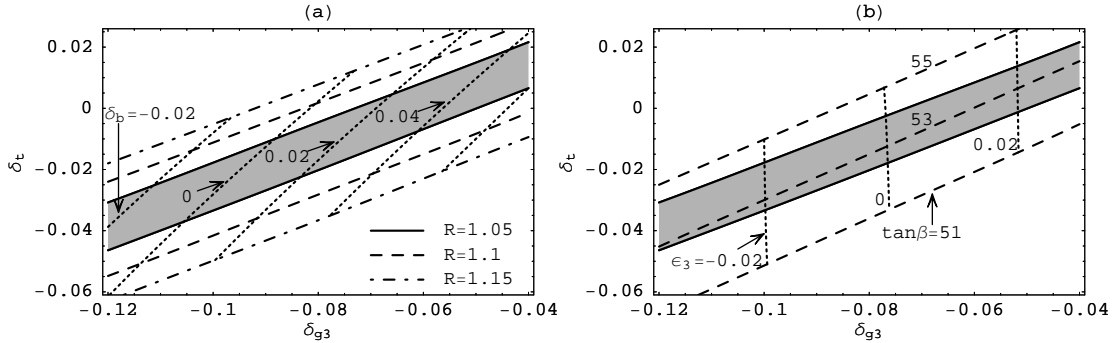} 
\vspace{-.5cm} 
\caption{Same as Fig. 1, but for $k_Y=23/21$ with vector-like charged singlets (one pair with
$Q=\pm1$ and one pair with $Q=\pm2/3$) added at $m_Z$.} \label{2321}
\end{figure}

Another way to restore gauge coupling unification while preserving Yukawa unification
is to add vector-like charged singlets and allowing fractional charges. As an example, we again take
$k_Y=23/21$, and add two pairs of charged singlets with mass $m_Z$ and charges $\pm1$ and
$\pm2/3$. As shown in Fig. \ref{2321}, gauge-Yukawa unification is then achieved similar to the
canonical case.

\begin{figure}[t] 
\centering
\includegraphics[angle=0, width=9cm]{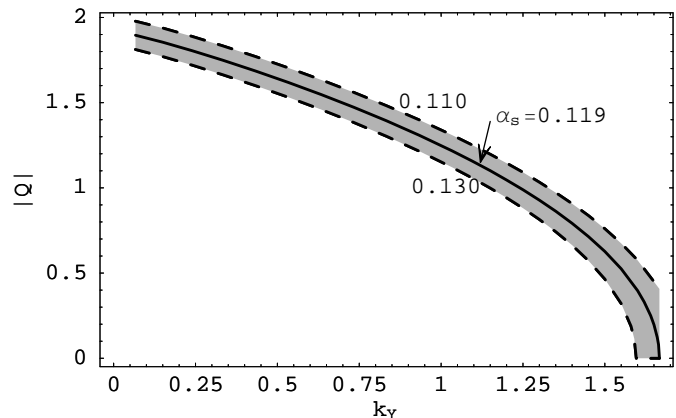} 
\vspace{-.5cm} 
\caption{$|Q|$ of the vector-like charged singlet with mass $m_Z$
allowing unification for $k_Y$ in the range $1/15$ to $5/3$.} \label{u1y}
\end{figure}

In Fig. \ref{u1y}, we show the charge of a vector-like charged singlet pair with mass $m_Z$
allowing unification, for $k_Y$ in the range $1/15$ to $5/3$. (Adding one pair
with charges $\pm Q$ is equivalent at one-loop to adding multiple pairs with charges $\pm Q_i$
if $Q^2=\sum_i Q^2_i$.) 
Here we choose $\delta_{t,b,\tau,g_i}$ such that $Q=0$ for $k_Y=5/3$ and $\alpha^{\rm MS}_s(m_Z)=0.119$.
The $\pm0.01$ uncertainty we display for $\alpha^{\rm MS}_s(m_Z)$ represents both SUSY and GUT threshold
corrections.

For fractionally charged singlets, there is a constraint on
particle per nucleon of about $10^{-22}$ \cite{Lee:2002sa}. This requires
the particle mass $M$ to be $\gtrsim10^4 T_r$, where $T_r$ is the reheating
temperature \cite{Kudo:2001ie}.\footnote{Since the fractionally charged
particle is not neutralized it may have difficulty getting past the
heliopause if it's not accelerated by astrophysical processes. This
may reduce the abundance on earth a few orders of magnitude, but since
the abundance is very sensitive to $M/T_r$, the conclusion does not
change much, and conservatively we can say $M\gtrsim10^3 T_r$.} Since $T_r$
\enlargethispage{\baselineskip}
can be as low as a few MeV, this in principle allows fractionally charged singlets
as light as allowed by accelerator searches.
The mass limit from accelerators is around $m_Z$ (for a review see Ref. \cite{Perl:2004qc}).

\section{Higgs Mass}
We end the paper with some remarks on the Higgs mass, where by the Higgs mass
we refer to the mass of the light $CP$-even scalar. Assuming that $m_Z\ll m_{\rm SUSY}$, where
$m_{\rm SUSY}$ is the characteristic supersymmetry particle mass scale, the theory below
$m_{\rm SUSY}$ is the SM with threshold effects at $m_{\rm SUSY}$.
The SM Higgs quartic coupling at $m_{\rm SUSY}$ is given by
\begin{equation}
\lambda=\frac{1}{4}(g_Y^2+g_2^2)\cos^2 2\beta=\frac{1}{4}\left(\frac{g_2^2}{\cos^2\theta_W}\right)\cos^2 2\beta\,,
\end{equation}
where $\tan\beta$ is the ratio of the two supersymmetric Higgs vacuum
expectation values, and $\theta_W$ is the Weinberg angle. Since 
$\cos^2\theta_W=k_Y/(1+k_Y)$ at $M_{\rm GUT}$, $\theta_W$ at $m_{\rm SUSY}$
depends on $k_Y$. The Higgs mass therefore also depends
on the value of $k_Y$, but for SUSY broken at the TeV scale the
effect is numerically insignificant, of order a few hundred
MeV. The Higgs mass predictions are therefore practically the same
as in canonical MSSM \cite{Carena:1995wu} and SUSY $SO(10)$ for the case with
third-family Yukawa unification \cite{Blazek:2001sb,Auto:2003ys}.  
The Higgs mass upper bound for $m_t=172.5$ GeV and $m_{\rm SUSY}=1$ TeV
is $\approx130$ GeV \cite{Carena:1995wu}.


\section{Conclusion}
We have considered a class of orbifold GUTs based on 6D ${\cal N} = (1,
1)$ and 7D ${\cal N} = 1$ supersymmetric $SU(N)$ gauge theories, where the 4D gauge group
is $SU(3)_C\times SU(2)_L\times U(1)_Y$  below the compactification scale.  For the $SU(5)$
model the only zero mode that can be introduced in the bulk is a quark doublet,
while the $SU(6)$ model allows gauge-Higgs unification. Finally, we can have
gauge-Yukawa unification for the third family in $SU(7)$ or
 higher rank groups.  Depending on the model, the $U(1)_Y$ normalization
factor $k_Y$ is either uniquely determined to have a non-canonical value or
lies in a continuous interval. Gauge coupling unification and gauge-Yukawa unification
can be obtained for non-canonical $k_Y$ values by adding particles to the MSSM 
spectrum. As examples, we introduce a minimal set of vector-like multiplets
with SM quantum numbers or fractionally charged color singlets, assuming masses in
the TeV range. The existence of such particles will be tested by the upcoming LHC.

\section*{Acknowledgments}

This work is supported in part by DOE Grant   \# DE-FG02-84ER40163
(I.G.), \#DE-FG02-96ER40959 (T.L.),   \# DE-FG02-91ER40626 (Q.S. and
V.N.S.), and by a University of Delaware graduate fellowship
(V.N.S.).


\section*{Appendix A: Six-Dimensional Orbifold Models}

We consider  6D space-time which can be factorized 
into a product of 4D Minkowski space-time $M^4$ and the torus $T^2$
which is homeomorphic to $S^1\times S^1$. The 6D
coordinates are $x^{\mu}$, ($\mu = 0, 1, 2, 3$),
$ x^5$ and $ x^6$. 
The radii for the circles along the $x^5 $  and $x^6 $ directions are
$R_1$ and $R_2$, respectively. We define the
complex coordinate
\begin{eqnarray}
z \equiv{1\over 2} \left(x^5 + i x^6\right)~,~\,
\end{eqnarray}
in which case
the torus $T^2$ can be defined as $C^1$ modulo the
equivalence classes: 
\begin{eqnarray}
z \sim z+ \pi R_1 ~,~ z \sim z +  \pi R_2 e^{{\rm i}\theta} ~.~\,
\end{eqnarray}

To define the orbifold $T^2/Z_6$, we require that $R_1=R_2\equiv R$
and $\theta = \pi/3$.
Then $T^2/Z_6$ orbifold is obtained from $T^2$ as:
\begin{eqnarray}
\Gamma_T:~~~z \sim \omega  z~,~\,
\end{eqnarray}
where $\omega =e^{{\rm i}\pi/3} $. There is one $Z_6$ fixed point:
$z=0$, two $Z_3$ fixed points: 
 $z=\pi R e^{{\rm i}\pi/6}/{\sqrt 3}$ and
$z=2 \pi R e^{{\rm i}\pi/6}/{\sqrt 3}$, and three $Z_2$ fixed points:
$z=\sqrt 3 \pi R e^{{\rm i}\pi/6}/2$, $z=\pi R/2$ and $z= \pi R e^{{\rm i}\pi/3}/2$.

The ${\cal N} = (1, 1)$ supersymmetry in 6D
has 16 supercharges and
 corresponds to  ${\cal N}=4$ supersymmetry in 4D,
so that only the gauge multiplet can be introduced in the bulk.  This
multiplet can be decomposed under  4D
 ${\cal N}=1$ supersymmetry into a vector
multiplet $V$ and three chiral multiplets $\Sigma_1$, $\Sigma_2$, and 
$\Sigma_3$ in the adjoint representation, where the fifth and sixth 
components of the gauge
field, $A_5$ and $A_6$ are contained in the lowest component of $\Sigma_1$.
The SM fermions can be on the 3-branes at the $Z_6$
fixed points. Here, we follow the conventions in Ref.~\cite{Li:2003ee}.

For the bulk gauge group $G$, we write down the  bulk action 
in the Wess-Zumino gauge and 4D ${\cal N}=1$ supersymmetry
language~\cite{NMASWS},
\begin{eqnarray}
  {\cal S} &=& \int d^6 x \Biggl\{
  {\rm Tr} \Biggl[ \int d^2\theta \left( \frac{1}{4 \kappa g^2} 
  {\cal W}^\alpha {\cal W}_\alpha + \frac{1}{\kappa g^2} 
  \left( \Sigma_3 \partial \Sigma_2   - \frac{1}{\sqrt{2}} \Sigma_1 
  [\Sigma_2, \Sigma_3] \right) \right) + {\rm h.c.} \Biggr] 
\nonumber\\
  && + \int d^4\theta \frac{1}{\kappa g^2} {\rm Tr} \Biggl[ 
  (\sqrt{2} \partial_z^\dagger + \Sigma_1^\dagger) e^{-V} 
  (-\sqrt{2} \partial_z + \Sigma_1) e^{V}
 + \partial_z^\dagger e^{-V} \partial_z e^{V} \Biggr]
\nonumber\\
&&+ \int d^4\theta \frac{1}{\kappa g^2} {\rm Tr} \Biggl[
   \Sigma_2^\dagger e^{-V} \Sigma_2  e^{V}
  + {\Sigma_3}^\dagger e^{-V} \Sigma_3 e^{V} 
\Biggr] \Biggr\}~,~\,
\label{eq:t2z6action}
\end{eqnarray}
where $\kappa$ is the normalization of the group generator,  and ~${\cal
W_{\alpha}}$~ denotes the gauge field strength. 
From the above action, we obtain  
the transformations of the vector multiplet 
\begin{eqnarray}
  V(x^{\mu}, ~\omega z, ~\omega^{-1} {\bar z}) &=& R_{\Gamma_T}
 V(x^{\mu}, ~z, ~{\bar z}) R_{\Gamma_T}^{-1}~,~\,
\label{Vtrans-6D}
\end{eqnarray}
\begin{eqnarray}
  \Sigma_1(x^{\mu}, ~\omega z, ~\omega^{-1} {\bar z}) &=& 
\omega^{-1} R_{\Gamma_T}
\Sigma_1(x^{\mu}, ~z, ~{\bar z}) R_{\Gamma_T}^{-1}~,~\,
\label{S1trans-6D}
\end{eqnarray}
\begin{eqnarray}
   \Sigma_2(x^{\mu}, ~\omega z, ~\omega^{-1} {\bar z}) &=& 
\omega^{-1-k} R_{\Gamma_T} 
\Sigma_2(x^{\mu}, ~z, ~{\bar z})  R_{\Gamma_T}^{-1}~,~\,
\label{S2trans-6D}
\end{eqnarray}
\begin{eqnarray}
 \Sigma_3(x^{\mu}, ~\omega z, ~\omega^{-1} {\bar z})  &=& 
\omega^{2+k} R_{\Gamma_T} 
\Sigma_3(x^{\mu}, ~z, ~{\bar z}) R_{\Gamma_T}^{-1}~,~\,
\label{S3trans-6D}
\end{eqnarray}
where $R_{\Gamma_T}$ is non-trivial
to break the bulk gauge group $G$. To preserve 4D ${\cal N}=1$ supersymmetry, we 
obtain $k=0$ or $k=1$~\cite{Li:2003ee}.

The $Z_6$ transformation properties for the decomposed components
of $V$, $\Sigma_1$, $\Sigma_2$,  and $\Sigma_3$ in the $SU(7)$ models are
\begin{equation}
V : \left(
\begin{array}{cccc}
1 & \omega^{-n_1} & \omega^{-n_2} & 
\omega^{-n_3}  \\
\omega^{n_1} & 1 & \omega^{n_1-n_2} &
\omega^{n_1-n_3}  \\
\omega^{n_2} & \omega^{n_2-n_1} & 1 & \omega^{n_2-n_3} \\
\omega^{n_3} & \omega^{n_3-n_1} & \omega^{n_3-n_2} & 1 
\end{array}
\right)  +  1 ~,~\,
\label{T2trans1}
\end{equation}
\begin{equation}
\Sigma_1 : \left(
\begin{array}{cccc}
\omega^{-1} & \omega^{-n_1-1} & \omega^{-n_2-1} & 
\omega^{-n_3-1} \\
\omega^{n_1-1} & \omega^{-1} & \omega^{n_1-n_2-1} &
\omega^{n_1-n_3-1} \\
\omega^{n_2-1} & \omega^{n_2-n_1-1} &  \omega^{-1}& \omega^{n_2-n_3-1} \\
\omega^{n_3-1} & \omega^{n_3-n_1-1} & \omega^{n_3-n_2-1} & \omega^{-1}  
\end{array}
\right)  +  \omega^{-1} ~,~\,
\label{T2trans2}
\end{equation}
\begin{equation}
\Sigma_2 : \left(
\begin{array}{cccc}
\omega^{-1-k} & \omega^{-n_1-1-k} & \omega^{-n_2-1-k} & 
\omega^{-n_3-1-k} \\
\omega^{n_1-1-k} & \omega^{-1-k} & \omega^{n_1-n_2-1-k} &
\omega^{n_1-n_3-1-k} \\
\omega^{n_2-1-k} & \omega^{n_2-n_1-1-k} &  \omega^{-1-k}& \omega^{n_2-n_3-1-k} \\
\omega^{n_3-1-k} & \omega^{n_3-n_1-1-k} & \omega^{n_3-n_2-1-k} & \omega^{-1-k} 
\end{array}
\right)  +  \omega^{-1-k} ~,~\,
\label{T2trans3}
\end{equation}
\begin{equation}
\Sigma_3 : \left(
\begin{array}{cccc}
\omega^{2+k} & \omega^{-n_1+2+k} & \omega^{-n_2+2+k} & 
\omega^{-n_3+2+k} \\
\omega^{n_1+2+k} & \omega^{2+k} & \omega^{n_1-n_2+2+k} &
\omega^{n_1-n_3+2+k} \\
\omega^{n_2+2+k} & \omega^{n_2-n_1+2+k} & \omega^{2+k} & \omega^{n_2-n_3+2+k} \\
\omega^{n_3+2+k} & \omega^{n_3-n_1+2+k} & \omega^{n_3-n_2+2+k} &
\omega^{2+k} 
\end{array}
\right)  +  \omega^{2+k} ~,~\,
\label{T2trans4}
\end{equation}
where the zero modes transform as $1$.
Note that $n_1 \not= n_2 \not= n_3 \not= 0$ and
from Eqs. (\ref{T2trans1})--(\ref{T2trans4}), 
 we obtain that
for the zero modes, the 6D 
${\cal N} = (1, 1) $ supersymmetric $SU(7)$ gauge symmetry is broken 
down to  4D ${\cal N}=1$ supersymmetric
$SU(3)_C\times SU(2)_L\times U(1)_Y \times 
 U(1)_{\beta} \times U(1)_{\gamma}$ gauge symmetry.



\section*{Appendix B: Seven-Dimensional Orbifold Models}

 We consider a 7D space-time $M^4\times T^2/Z_6 \times
S^1/Z_2$ with coordinates $x^{\mu}$, ($\mu = 0, 1, 2, 3$), $x^5$,
$x^6$ and $x^7$. The torus $T^2$ is homeomorphic to $S^1\times S^1$
and  the radii of the circles along the $x^5$, $x^6$ and $x^7$
directions are $R_1$, $R_2$, and $R'$, respectively. We introduce a
complex coordinate $z$ for $T^2$ and a real coordinate $y$ for
$S^1$,
\begin{eqnarray}
z \equiv{1\over 2} \left(x^5 + i x^6\right),~~~~~~~~~~ y \equiv x^7.
\end{eqnarray}
The orbifold $T^2/Z_6$ has been defined in  Appendix A, while 
the orbifold $S^1/Z_2$  is obtained from
$S^1$ by moduloing the equivalent class
\begin{eqnarray}
\Gamma_S:~~~y\sim -y~.~\,
\end{eqnarray}
There are two fixed points: $y=0$ and $y=\pi R'$.

The 7D ${\cal N}=1$
supersymmetry has 16 supercharges corresponding to ${\cal
N}=4$ supersymmetry in 4D, and only the gauge multiplet can be
introduced in the bulk.  This multiplet can be decomposed under  4D
 ${\cal N}=1$ supersymmetry into a gauge vector
multiplet $V$ and three chiral multiplets $\Sigma_1$, $\Sigma_2$,
and $\Sigma_3$, all in the adjoint representation, where the fifth and
sixth components of the gauge field, $A_5$ and $A_6$, are contained
in the lowest component of $\Sigma_1$, and the seventh component of
the gauge field $A_7$ is contained in the lowest component of
$\Sigma_2$.

We express the  bulk action in the Wess--Zumino gauge and 4D ${\cal
N}=1$ supersymmetry notation~\cite{NMASWS}
\begin{eqnarray}
  {\cal S} &=& \int d^7 x \Biggl\{
  {\rm Tr} \Biggl[ \int d^2\theta \left( \frac{1}{4 \kappa g^2}
  {\cal W}^\alpha {\cal W}_\alpha + \frac{1}{\kappa g^2}
  \left( \Sigma_3 \partial_z \Sigma_2 + \Sigma_1 \partial_y \Sigma_3
   - \frac{1}{\sqrt{2}} \Sigma_1
  [\Sigma_2, \Sigma_3] \right) \right)
\nonumber\\
  &&
+ h.c. \Biggr]
 + \int d^4\theta \frac{1}{\kappa g^2} {\rm Tr} \Biggl[
  (\sqrt{2} \partial_z^\dagger + \Sigma_1^\dagger) e^{-V}
  (-\sqrt{2} \partial_z + \Sigma_1) e^{V}
 + \partial_z^\dagger e^{-V} \partial_z e^{V}
\nonumber\\
  && +
  (\sqrt{2} \partial_y + \Sigma_2^\dagger) e^{-V}
  (-\sqrt{2} \partial_y + \Sigma_2) e^{V}
 + \partial_y e^{-V} \partial_y e^{V}
+ {\Sigma_3}^\dagger e^{-V} \Sigma_3 e^{V} \Biggr] \Biggr\}.~\,
\label{action7}
\end{eqnarray}
 From the above
action, we obtain the transformations of the vector multiplet:
\begin{eqnarray}
  V(x^{\mu}, ~\omega z, ~\omega^{-1} {\bar z},~y) &=& R_{\Gamma_T}
 V(x^{\mu}, ~z, ~{\bar z},~y) R_{\Gamma_T}^{-1}~,~\,
\label{TVtrans}
\end{eqnarray}
\begin{eqnarray}
  \Sigma_1(x^{\mu}, ~\omega z, ~\omega^{-1} {\bar z},~y) &=&
\omega^{-1} R_{\Gamma_T} \Sigma_1(x^{\mu}, ~z, ~{\bar z},~y)
R_{\Gamma_T}^{-1}~,~\, \label{T1trans}
\end{eqnarray}
\begin{eqnarray}
   \Sigma_2(x^{\mu}, ~\omega z, ~\omega^{-1} {\bar z},~y) &=&
 R_{\Gamma_T}
\Sigma_2(x^{\mu}, ~z, ~{\bar z},~y)  R_{\Gamma_T}^{-1}~,~\,
\label{T2trans}
\end{eqnarray}
\begin{eqnarray}
 \Sigma_3(x^{\mu}, ~\omega z, ~\omega^{-1} {\bar z},~y)  &=&
\omega R_{\Gamma_T} \Sigma_3(x^{\mu}, ~z, ~{\bar z},~y)
R_{\Gamma_T}^{-1}~,~\, \label{T3trans}
\end{eqnarray}
\begin{eqnarray}
  V(x^{\mu}, ~z, ~ {\bar z},~-y) &=& R_{\Gamma_S}
 V(x^{\mu}, ~z, ~{\bar z},~y) R_{\Gamma_S}^{-1}~,~\,
\label{SVtrans}
\end{eqnarray}
\begin{eqnarray}
  \Sigma_1(x^{\mu}, ~ z, ~ {\bar z},~-y) &=&
 R_{\Gamma_S}
\Sigma_1(x^{\mu}, ~z, ~{\bar z},~y) R_{\Gamma_S}^{-1}~,~\,
\label{S1trans}
\end{eqnarray}
\begin{eqnarray}
   \Sigma_2(x^{\mu}, ~ z, ~ {\bar z},~-y) &=&
-  R_{\Gamma_S} \Sigma_2(x^{\mu}, ~z, ~{\bar z},~y)
R_{\Gamma_S}^{-1}~,~\, \label{S2trans}
\end{eqnarray}
\begin{eqnarray}
 \Sigma_3(x^{\mu}, ~ z, ~ {\bar z},~-y)  &=&
- R_{\Gamma_S} \Sigma_3(x^{\mu}, ~z, ~{\bar z},~y)
R_{\Gamma_S}^{-1}~,~\, \label{S3trans}
\end{eqnarray}
where we introduce  non-trivial transformations $R_{\Gamma_T}$ and
$R_{\Gamma_S}$ to break the bulk gauge group $G$.




The $Z_6\times Z_2$ transformation properties for the decomposed
components of $V$, $\Sigma_1$, $\Sigma_2$,  and $\Sigma_3$
in  the $SU(8)$ model are given by

\begin{equation}
V : \left(
\begin{array}{ccccc}
(1, +) & (\omega^{-n_1}, +) & (\omega^{-n_1}, -) & 
(1, -) & (\omega^{-n_2}, +) \\
(\omega^{n_1}, +) & (1, +) & (1, -) & (\omega^{n_1}, -)
& (\omega^{n_1-n_2}, +) \\
(\omega^{n_1}, -) & (1, -) & (1, +) & (\omega^{n_1}, +)
& (\omega^{n_1-n_2}, -) \\
(1, -) & (\omega^{-n_1}, -) & (\omega^{-n_1}, +)
& (1, +) & (\omega^{-n_2}, -) \\
(\omega^{n_2}, +) & (\omega^{n_2-n_1}, +) &
 (\omega^{n_2-n_1}, -) & (\omega^{n_2}, -) & (1, +) 
\end{array}
\right)  +  (1, +) ~,~\,
\label{SU8-GTBT-trans-1}
\end{equation}

\begin{equation}
\Sigma_1 : \left(
\begin{array}{ccccc}
(\omega^{-1}, +) & (\omega^{-n_1-1}, +) & (\omega^{-n_1-1}, -) & 
(\omega^{-1}, -) & (\omega^{-n_2-1}, +) \\
(\omega^{n_1-1}, +) & (\omega^{-1}, +) & (\omega^{-1}, -) & (\omega^{n_1-1}, -)
& (\omega^{n_1-n_2-1}, +) \\
(\omega^{n_1-1}, -) & (\omega^{-1}, -) & (\omega^{-1}, +) & (\omega^{n_1-1}, +)
& (\omega^{n_1-n_2-1}, -) \\
(\omega^{-1}, -) & (\omega^{-n_1-1}, -) & (\omega^{-n_1-1}, +)
& (\omega^{-1}, +) & (\omega^{-n_2-1}, -) \\
(\omega^{n_2-1}, +) & (\omega^{n_2-n_1-1}, +) &
 (\omega^{n_2-n_1-1}, -) & (\omega^{n_2-1}, -) & (\omega^{-1}, +) 
\end{array}
\right) +  (\omega^{-1}, +) ~,~\,
\label{SU8-GTBT-trans-2}
\end{equation}

\begin{equation}
\Sigma_2 : \left(
\begin{array}{ccccc}
(1, -) & (\omega^{-n_1}, -) & (\omega^{-n_1}, +) & 
(1, +) & (\omega^{-n_2}, -) \\
(\omega^{n_1}, -) & (1, -) & (1, +) & (\omega^{n_1}, +)
& (\omega^{n_1-n_2}, -) \\
(\omega^{n_1}, +) & (1, +) & (1, -) & (\omega^{n_1}, -)
& (\omega^{n_1-n_2}, +) \\
(1, +) & (\omega^{-n_1}, +) & (\omega^{-n_1}, -)
& (1, -) & (\omega^{-n_2}, +) \\
(\omega^{n_2}, -) & (\omega^{n_2-n_1}, -) &
 (\omega^{n_2-n_1}, +) & (\omega^{n_2}, +) & (1, -) 
\end{array}
\right) +  (1, -) ~,~\,
\label{SU8-GTBT-trans-3}
\end{equation}

\begin{equation}
\Sigma_3 : \left(
\begin{array}{ccccc}
(\omega, -) & (\omega^{-n_1+1}, -) & (\omega^{-n_1+1}, +) & 
(\omega, +) & (\omega^{-n_2+1}, -) \\
(\omega^{n_1+1}, -) & (\omega, -) & (\omega, +) & (\omega^{n_1+1}, +)
& (\omega^{n_1-n_2+1}, -) \\
(\omega^{n_1+1}, +) & (\omega, +) & (\omega, -) & (\omega^{n_1+1}, -)
& (\omega^{n_1-n_2+1}, +) \\
(\omega, +) & (\omega^{-n_1+1}, +) & (\omega^{-n_1+1}, -)
& (\omega, -) & (\omega^{-n_2+1}, +) \\
(\omega^{n_2+1}, -) & (\omega^{n_2-n_1+1}, -) &
 (\omega^{n_2-n_1+1}, +) & (\omega^{n_2+1}, +) & (\omega, -) 
\end{array}
\right) +  (\omega, -) ~,~\,
\label{SU8-GTBT-trans-4}
\end{equation}
From Eqs. (\ref{SU8-GTBT-trans-1})--(\ref{SU8-GTBT-trans-4}), 
we obtain that  the 7D
${\cal N} = 1 $ supersymmetric $SU(8)$ gauge symmetry is broken 
down to  4D ${\cal N}=1$ supersymmetric
$SU(3)_C\times SU(2)_L\times U(1)_Y \times U(1)_{\alpha}
\times U(1)_{\beta} \times U(1)_{\gamma}$ gauge 
symmetry~\cite{Li:2001tx}.



\end{document}